\documentclass[aps,prl,preprint,groupedaddress,floatfix]{revtex4}

\usepackage{graphicx,latexsym,bm}

\begin{document}


\title{A thermodynamic model for the melting of supported metal nanoparticles}


\author{S. C. Hendy}
\email{s.hendy@irl.cri.nz}
\affiliation{Industrial Research Ltd, Lower Hutt, New Zealand}
\affiliation{MacDiarmid Institute for Advanced Materials
and Nanotechnology, School of Chemical and Physical Sciences, Victoria University of Wellington, New Zealand}


\date{\today}

\begin{abstract}
We construct a simple thermodynamic model to describe the melting of a supported metal nanoparticle with a spherically curved free surface both with and without surface melting. We use the model to investigate the results of recent molecular dynamics simulations, which suggest the melting temperature of a supported metal particle is the same as that of a free spherical particle with the same surface curvature. Our model shows that this is only the case when the contact angles of the supported solid and liquid particles are similar. This is also the case for the temperature at which surface melting begins.   
\end{abstract}

\pacs{}

\maketitle

\section{Introduction}

Despite decades of study, the melting of nanoparticles continues to generate interest \cite{Haberland02, Haberland05, Breaux05, Hendy06}. In general, the melting temperature of  spherical nanoparticles has been found to decrease in proportion to the surface area to volume ratio of the particle \cite{Buffat76}, as the surface free energy of a molten droplet is less than that of the corresponding solid particle. Although free nanoparticle calorimetry has advanced considerably in recent years \cite{Haberland01,Breaux03}, most experimental determinations of the melting points of nanoparticles are conducted with supported particles (gold \cite{Castro90}, lead \cite{BenDavid95} and tin \cite{Lai96}, for example). The melting of supported nanoparticles is also important in carbon nanotube growth and other catalytic processes \cite{Kuang00,Ding04}, and for the stability of devices assembled from nanoparticles \cite{partridge04,reichel06}. Thus it is of interest to study the effect of the substrate on the melting point of a supported nanoparticle.

Recent molecular dynamics simulations \cite{Ding06} of supported iron nanoparticles with a strongly interacting substrate found that the melting point of the particles was reduced in inverse proportion to the equilibrium surface curvature that results as they relax to wet the substrate. This statement also holds in the free particle limit since the curvature of a free spherical particle of radius $a$ is proportional to its surface to volume ratio, $3/a$. Interestingly, the simulations in Ref.~\cite{Ding06} found that the constant of proportionality between the decrease in melting point and the surface curvature did not depend on whether the particle was supported or free. In other words, the melting temperature of a supported particle that has a free surface with radius of curvature $a$, was found to be the same as that of a free spherical particle with the same surface curvature. The simulated nanoparticles in Ref.~\cite{Ding06} also exhibited surface melting prior to complete melting. Surface melting is phenomena thought to occur both on certain bulk metal surfaces \cite{vanderVeen90} and in certain metal nanoparticles \cite{Bachels00, Tosatti05}. 

In this paper we use a simple thermodynamic model to investigate the role of the substrate in both melting and surface melting of metal nanoparticles. Our model suggests that the result in Ref.~\cite{Ding06}, that the relative decrease in melting point is proportional to the solid particle surface curvature, only holds when the contact angles of the supported solid and liquid particles with the substrate are similar. We also show that supported clusters will exhibit surface melting under certain circumstances, and that the surface melting temperature in free and supported particles in clusters with same surface curvature is the same only when the contact angles of the supported solid and liquid phases coincide.

\section{Geometry of supported particles}

We start by considering a solid nanoparticle, initially spherical with radius $a$, that is placed on a flat substrate. We neglect the effect of faceting, curvature dependent surface energies and internal strains due to epitaxial mismatch with the substrate. Furthermore, we will assume that the particle has relaxed to its equilibrium geometry i.e. that the nanoparticle has relaxed to ''wet" the substrate. Provided the nanoparticle is heated sufficiently slowly, the particle should relax to this geometry prior to melting. With these simplifying assumptions, the geometry of the relaxed particle can be approximated by a spherical cap, as shown in Fig.~\ref{fig:1}, with dimensions parameterised by either the cap height $H$, or radius of curvature of the free surface $R$, which minimizes the surface energy of the nanoparticle and substrate. 

The surface energy $\Gamma$ of the system can be written as:
\begin{equation}
\Gamma = 2 \pi R H \gamma_s  + \pi H (2 R - H) (\gamma_{sb} - \gamma_b) + \Gamma_b
\end{equation}
where $\gamma_s$ is the surface energy density of the free particle surface, $\gamma_b$ is the surface energy density of the substrate, $\gamma_{sb}$ is the particle-substrate interfacial energy density and $\Gamma_b$ is the total energy of the bare substrate. We will assume that the density of the particle $\rho_s$ does not depend on the contact angle so that the volume of the supported particle remains the same as that of the free particle.

Writing the volume of the particle as a function of H and R, it is then straightforward to show that $\Gamma$ is minimized if $H = -\left( \Delta \gamma_{sb} / \gamma_s \right) R$ where $\Delta \gamma_{sb} = \gamma_b - \gamma_s - \gamma_{sb}$. We note that $\Delta \gamma_{sb}$ is often called the spreading parameter in the context of wetting phenomena: if $\Delta \gamma_{sb} > 0$ then the particle will relax to fully wet the substrate. Here we are interested in the case where the particle does not fully wet the substrate (contact angles greater than zero) i.e. when $\Delta \gamma_{sb} < 0$ and $H/R = -\Delta \gamma_{sb} / \gamma_s >0$ at equilibrium. In fact this minimum value of $\Gamma$ can be written as
\begin{equation}
\label{eq:2}
\Gamma^* = 2 \pi \gamma_s \left( a \over R^*_s \right) a^2 + \Gamma_b,
\end{equation}
where $R^*_s$ is the corresponding radius of curvature of the supported solid nanoparticle, given by
\begin{equation}
\label{eq:3}
R^*_s=\left( \frac{2\gamma_s}{\Delta \gamma_{sb}} \right)^{2/3} \frac{a}{\left(3+\Delta \gamma_{sb}/\gamma_s \right)^{1/3}}.
\end{equation}
Thus $\Gamma^*$ and $R^*_s$ are the equilibrium surface energy and radius of curvature of the particle respectively. Note that the contact angle of the particle can range from 0 to 180 degrees depending on the value of the spreading parameter $\Delta \gamma_{sb}$. 

\section{Melting and surface melting}
 
In what follows we will assume that the density of the solid and liquid phases are identical i.e. $\rho_s = \rho_l = \rho$. We first consider the situation in which there is no surface melting. In this case, melting will occur at a temperature when the free energy of the solid particle wetting the substrate is equal to that of the corresponding liquid droplet wetting the substrate. If $\gamma_l$ is the surface energy density of the free liquid droplet and $R^*_l$ is the corresponding equilibrium radius of curvature, then the difference in free energy between the solid and liquid is     
\begin{equation}
\label{eq:4}
\Delta F = \frac{4}{3} \pi a^3 \rho \left(f_s - f_l + 3\left(\frac{\gamma_s}{R^*_s} - \frac{\gamma_l}{R^*_l} \right) \right)
\end{equation}
where $f_s$ ($f_l$) is the bulk free energy density of the solid (liquid). Now, using $f_l-f_s = \rho L \left( 1 - T/T_c \right)$, where $L$ is the latent heat of melting and $T_c$ is the bulk melting temperature, we find the melting temperature $T_m$ of the supported particle is given by:
\begin{eqnarray}
T_m & = & T_c \left( 1 - \frac{3}{\rho L}\left(\frac{\gamma_{s}}{R^*_s}- \frac{\gamma_{l}}{R^*_l} \right) \right) \nonumber \\
\label{eq:5} & = & T^{free}_m \left( R^*_s \right) + \frac{3 \gamma_l}{\rho R^*_s L} \left( 1- \frac{R^*_s}{R^*_l} \right) T_c 
\end{eqnarray}
Thus if $R^*_s = R^*_l = R^*$ then we recover the result of Ref.~\cite{Ding06}, namely that $T_m = T^{free}_m \left( R^* \right)$. In other words, if the contact angles of the solid and liquid droplets are equal, the melting temperature of the supported particle is the same as that of a free particle with an identical surface curvature, $a=R^*$. However, if the curvature of the supported liquid particle is different from that of the supported solid particle, it can be seen that the melting temperature will deviate from that found in Ref.~\cite{Ding06}.

\begin{figure}
\resizebox{\columnwidth}{!}{\includegraphics{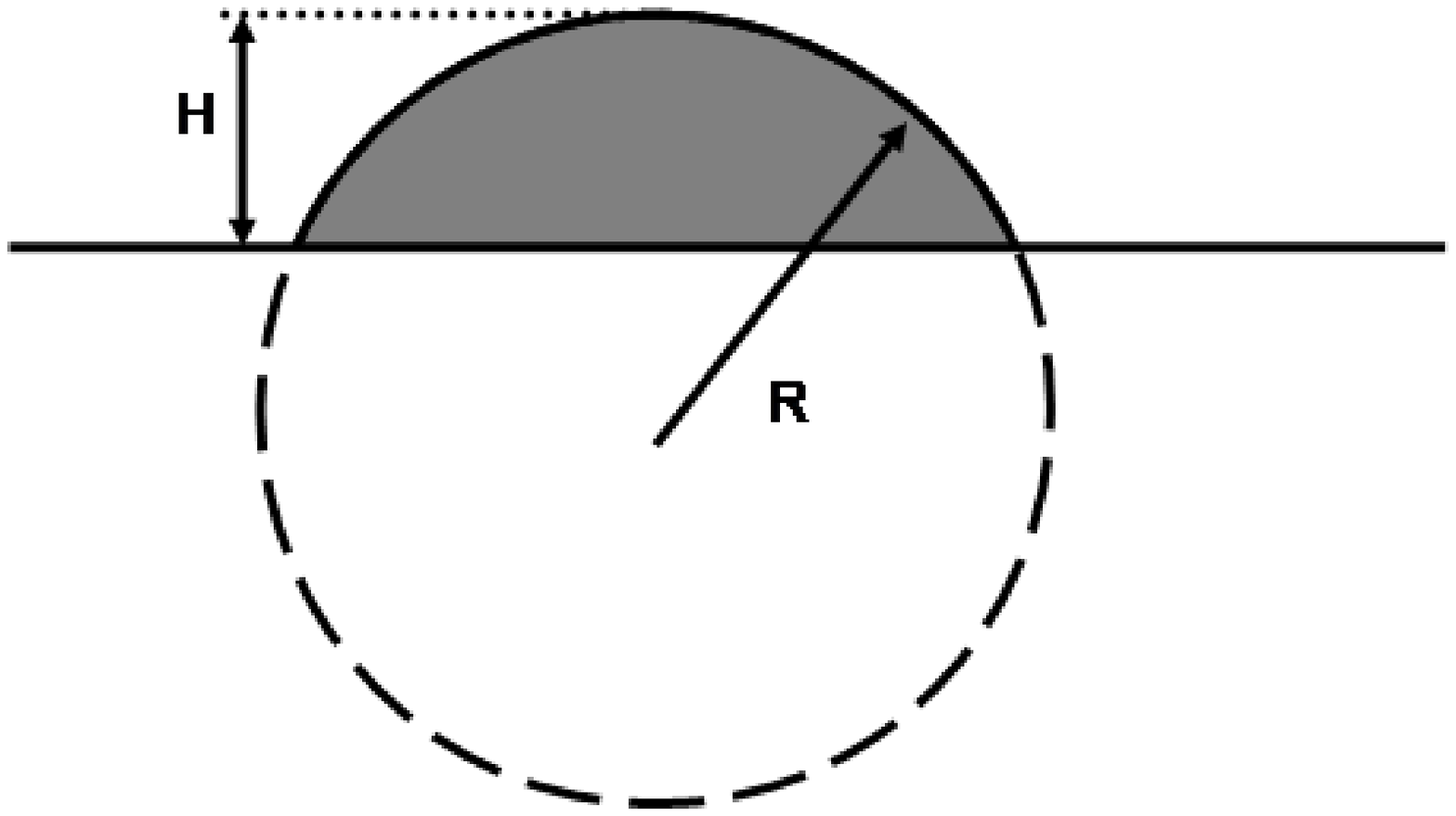}\includegraphics{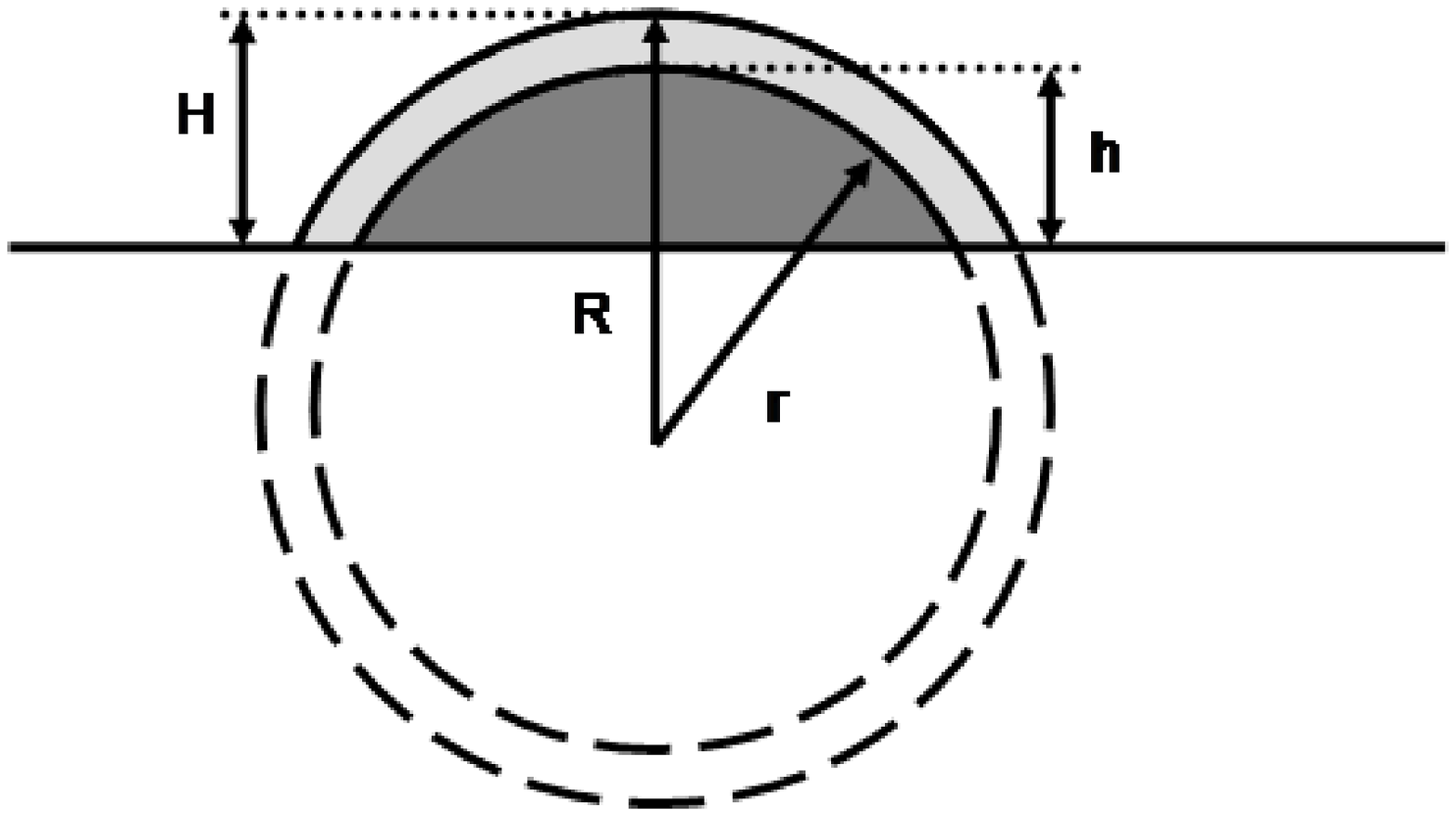}}%
\caption{\label{fig:1} The model for the geometry of a supported nanoparticle in equilibrium. We assume that the particle is a spherical cap of height $H$ and radius of curvature $R$ (left - the dashed lines simply illustrate the radius of curvature). At the onset of surface melting, we assume that the geometry is close to that of the solid particle in its equilibrium geometry and that that the solid particle (radius of curvature $r$ and height $h$) is initially wet by a molten layer of uniform thickness $\delta = R-r = H-h$ (right).}
\end{figure}

Now we consider surface melting as illustrated in figure~\ref{fig:1} which is thought to occur in many metals prior to melting \cite{vanderVeen90}. We are interested in determining the onset of melting, when the solid particle is wet by a thin layer of melt (thickness $\delta$) at the solid-vapor interface. We will assume that this melt forms a layer of uniform thickness with a geometry like that represented in figure~\ref{fig:1} with $\delta = R - r = H - h$. The total free energy of the surface melted particle is then a function of $\delta$: $F(\delta) = V_s(\delta)(f_s-f_l)+ V f_l + \Gamma (\delta)$ where $V_{s(l)}$ is the volume of the solid (liquid) and $\Gamma$ is the thickness dependent surface energy. In particular 
\[ \Gamma = \pi \left( 2 R H \gamma_l + r (2r-h) \gamma_{sb} + \delta (2 R - \delta) \gamma_{lb} + 2 r h \gamma_{sl}(\delta)\right) \]
where $\gamma_{sl}(\delta)=\gamma_{sl}+\Delta \gamma_{sl} \exp{\left(-\delta/\xi\right)}$ and $\xi$ is a correlation length that describes the thickness dependence of the interfacial energy in thin metallic liquid films \cite{vanderVeen90} (in Pb, for example, $\xi$ has been measured to be $\sim 0.6$ nm \cite{vanderVeen89}). As the surface melting proceeds, the curvature of the particle will relax to minimize the free energy i.e $R^*=R^*(\delta)$ where $R^*$ minimizes the free energy $F$ for a given $\delta$.    

In an isolated spherical nanoparticle of radius $a$, by minimizing the free energy $F(\delta)$ with respect to $\delta$ and setting $\delta=0$, one can show that surface melting begins at a temperature, $T_s(a)$ given by
\begin{equation}
\label{eq:6}
T^{free}_s (a)= T_c \left( 1 -  \frac{\Delta \gamma_{sl}}{\rho \xi L} + 2\frac{\left( \gamma_{s}-\gamma_{l} \right)}{\rho a L}\right).
\end{equation}
provided $\Delta \gamma_{sl} > 0$ and $a > \xi (\gamma_s-\gamma_l)/\Delta \gamma_{sl}$ (if $a$ is less than this, full melting will precede surface melting i.e. $T^{free}_s > T^{free}_m$ \cite{Bachels00}, and equation (\ref{eq:5}) will hold). 

For surface melting to occur in a supported solid nanoparticle with equilibrium curvature $R_s$, a minimum in the free energy $F(\delta)$ must appear at $\delta=0$. It is straightforward to show that a minimum in $F$ at $\delta=0$ occurs at the temperature $T_s$:
\begin{eqnarray}
T_s \left( R_s \right) & = & T^{free}_s \left( R_s \right) + \frac{T_c}{\rho R_s L} \left( \frac{\gamma_{s}\Delta \gamma_{lb}-\gamma_{l} \Delta \gamma_{sb}}{\Delta \gamma_{sb}} \right) \nonumber \\ 
\label{eq:7}
& = & T^{free}_s \left( R_s \right) +  \frac{\gamma_l T_c}{\rho R_s L} \left( \frac{\cos \theta_s - \cos \theta_l}{1-\cos \theta_s} \right) 
\end{eqnarray}
where $\theta_s$ and $\theta_l$ are the contact angles for solid particle and liquid particle respectively (defined via Young's relation $\gamma_{s(l)} \cos \theta_{s(l)} = \gamma_b - \gamma_{s(l)b}$). Once again, if the contact angles of the solid and liquid droplets are equal, then the temperature at which surface melting occurs is identical to that of a free particle with the same surface curvature, $R_s$ i.e. $T_s = T^{free}_s \left( R_s \right)$. Further, if $\cos \theta_s > \cos \theta_l$, so that the substrate favors contact with the solid over that with the liquid, the corresponding $T_s$ increases and vice versa. 

\begin{figure*}
\resizebox{\textwidth}{!}{\includegraphics{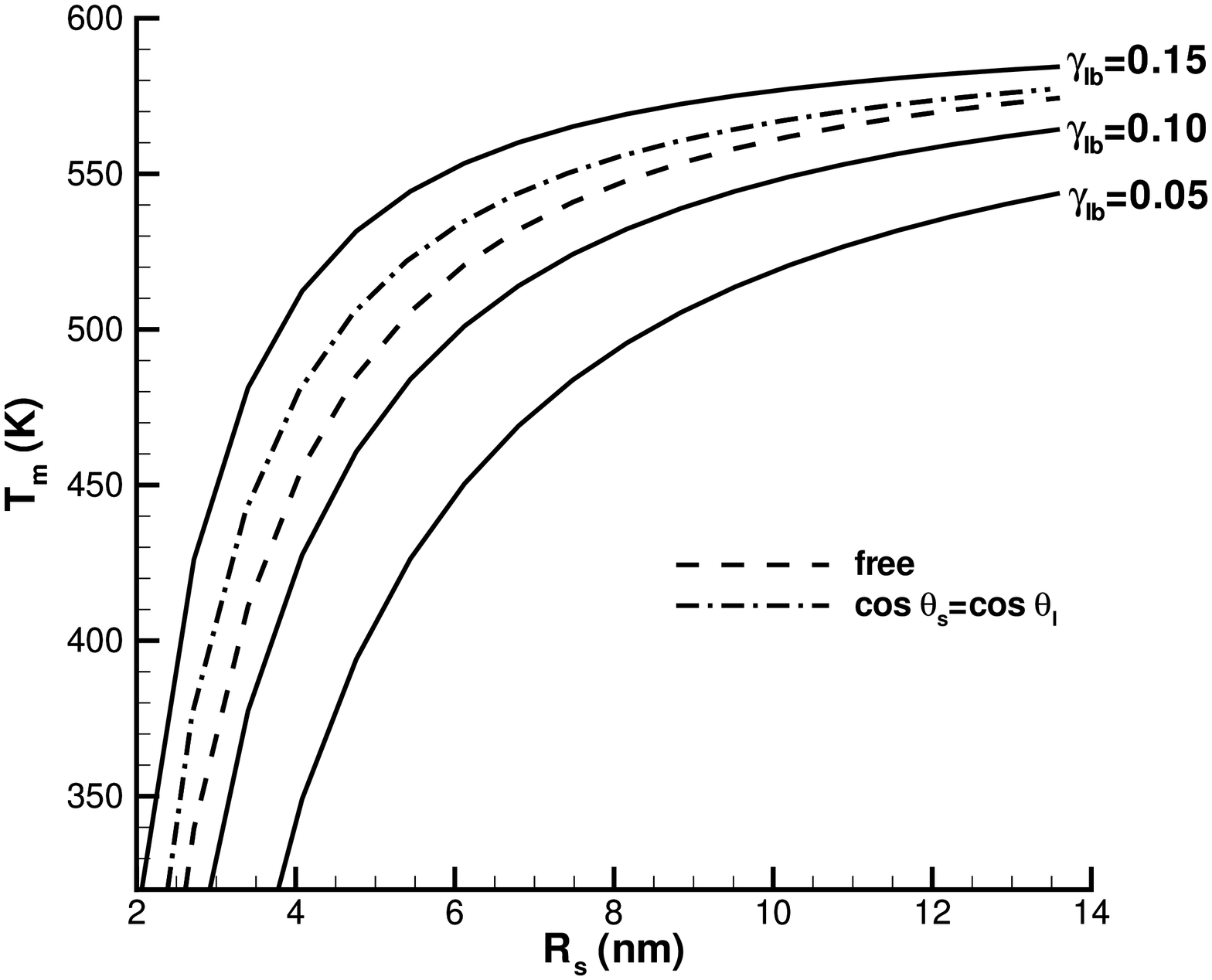} \includegraphics{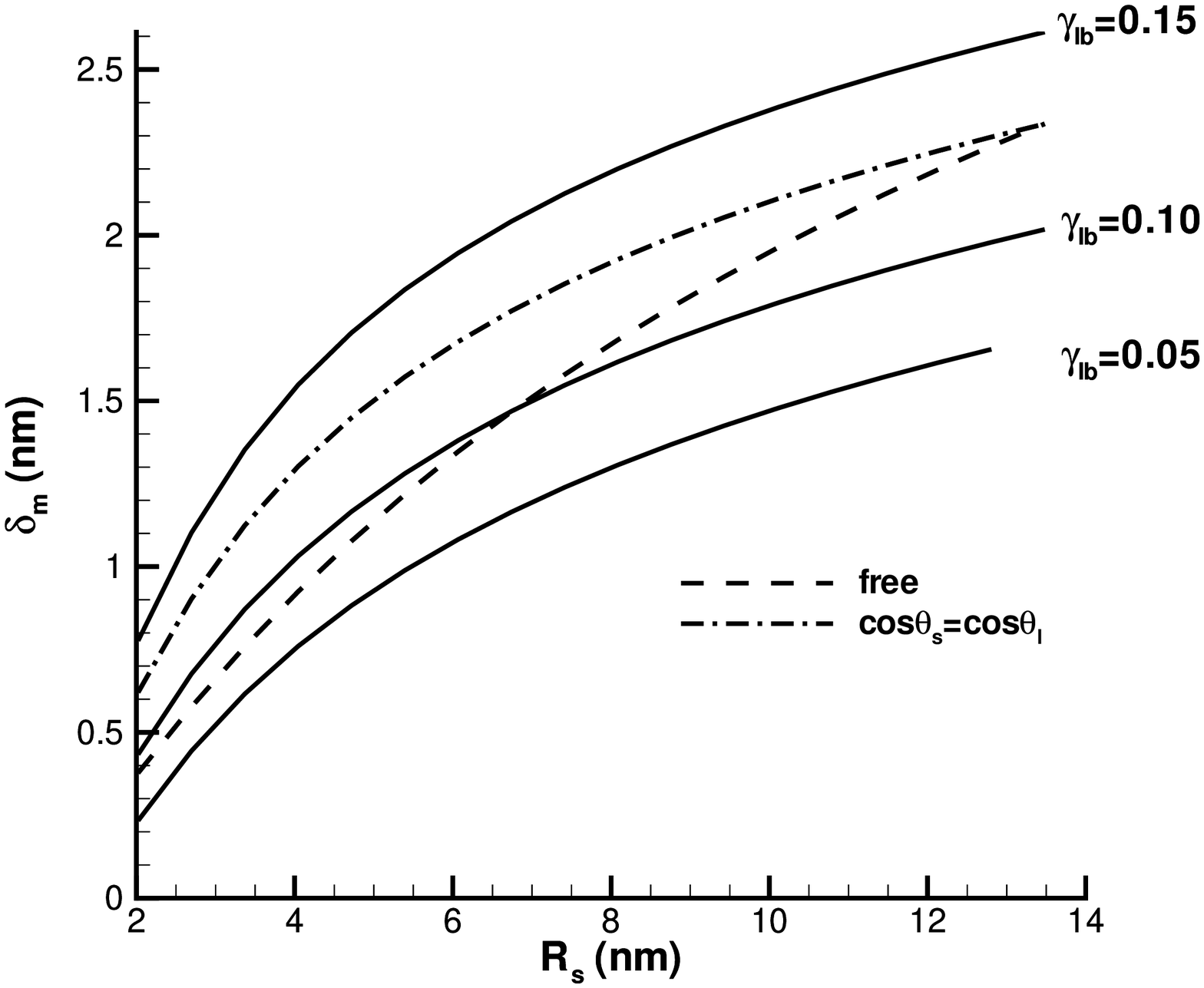}}
\caption{\label{fig:2} The melting temperature $T_m$ and critical liquid film thickness $\delta_m$ for supported Pb clusters as a function of the radius of curvature $R_s$ of the relaxed solid particle for $\gamma_{lb} = 0.05$, $0.10$, $0.15$ J m$^{-2}$ and in the case where $\cos \theta_{s} = \cos \theta_{l}$ ($\gamma_{lb} \simeq 0.13$ J m$^{-2}$). Also shown is the melting temperature of a free particle with radius $R_s$. Other surface energies used are $\gamma_{sv} = 0.61$, $\gamma_{lv} = 0.48$, $\gamma_{sl} = 0.05$, $\gamma_{b} = 0.25$ and $\gamma_{sb} = 0.1$ J m$^{-2}$ giving a contact angle of 75.8$^o$ for the solid supported cluster, and contact angles for the liquid droplets ranging from 78.0$^o$ to 65.4$^o$ respectively. Other parameters used were $\xi = 0.63$ nm, $\rho = 10950$ kg m$^{-3}$, $L = 22930$ J kg$^{-1}$ and $T_c = 600.65$ K \cite{BenDavid95}.}  
\end{figure*}

Complete melting will occur once the free energy of the surface melted particle, $F(\delta)$, equals that of the corresponding liquid droplet, $F_l$  i.e. at the temperature $T_m$ and liquid film thickness $\delta_m$ which satisfy $F(\delta_m)=F_l$. It is not possible to obtain an analytic expression for $\delta_m$ or $T_m$, but numerical solutions to the resulting equations are shown in figure~\ref{fig:2} as a function of $R_s$ for Pb particles. The figures clearly show the strong dependence of the melting temperature on the liquid droplet contact angle: a difference of $\sim 10^o$ in the molten particle contact angle can shift the melting point by $\sim 100$ K. Note that the melting point of a free particle with radius $R_s$ no longer coincides with that of a supported particle with radius of curvature $R_s$ when $\cos \theta_s = \cos \theta_l$, as in general the radius of curvature of the critical surface melted droplet will not be that of the solid particle (although the curves lie close to each other).  

\section{Conclusion}

We conclude that the melting temperature (and surface melting temperature, if the particle exhibits surface melting) of supported nanoparticles depends on the radius of curvature (or the contact angle) of both the supported solid and liquid particles. In general, we do not expect these curvatures to be the same: on a non-ideal solid substrate for example, epitaxial effects may favor one phase over the other. It is likely that the ideal substrate used in Ref~\cite{Ding06} resulted in very similar solid and liquid particle contact angles. We have shown that is only in this "ideal" case that the melting temperature of free and supported particles with the same curvature is coincident, whether they exhibit surface melting or otherwise. Thus, results from free particle melting, where the curvature of the solid and liquid particles remain substantially the same, have only limited applicability to supported particle melting.

\clearpage

\thispagestyle{empty}
\clearpage

\thispagestyle{empty}









\end{document}